\documentclass[12pt]{article}

\usepackage{amsmath}
\usepackage{latexsym}
\newcommand\be{\begin{equation}}
\newcommand\ee{\end{equation}}
\newcommand\bea{\begin{eqnarray}}
\newcommand\eea{\end{eqnarray}}
\newcommand\beas{\begin{eqnarray*}}
\newcommand\eeas{\end{eqnarray*}}

\def\tr{{\rm Tr}}

\begin{document}
\title{Laplacians in polar matrix coordinates and radial fermionization in higher dimensions}

\author{\\
Mthokozisi Masuku and Jo\~ao P. Rodrigues\footnote{Email: joao.rodrigues@wits.ac.za} \\
\\
National Institute for Theoretical Physics \\
School of Physics and Centre for Theoretical Physics \\
University of the Witwatersrand, Johannesburg\\
Wits 2050, South Africa 
}

\maketitle

\begin{abstract}
We consider the quantum mechanical hamiltonian of two, space indexed, hermitean matrices. By introducing matrix valued polar coordinates, we obtain the form of the laplacian acting on invariant states. For potentials depending only on the eigenvalues of the radial matrix, we establish that the radially invariant sector is equivalent to a system of non interacting $2+1$ dimensional fermions, and obtain its density description. For a larger number of matrices, the presence of a repulsive radial inter-eigenvalue potential is identified.     
\end{abstract}

\newpage

\noindent
\section{Introduction}

The study of multi-matrix models\footnote{By matrix models we mean integrals over matrices or the quantum mechanics of matrix valued degrees of freedom}, 
and particularly their large $N$ limit \cite{'t Hooft:1973jz}, is of great interest. 
It is well known, for instance, that the large $N$ limit of their description of D$0$ branes \cite{Polchinski:1995mt} has been conjectured to provide a definition of $M$ theory \cite{Banks:1996vh}. 
In the context of the AdS/CFT duality \cite{Maldacena:1997re}, \cite{Gubser:1998bc}, \cite{Witten:1998qj},  
due to supersymmetry and conformal invariance, correlators of supergravity and $1/2$ BPS states reduce to calculation of free matrix model overlaps 
\cite{Lee:1998bxa}, \cite{Corley:2001zk} or consideration of related matrix hamiltonians \cite{Berenstein:2004kk}. 
For stringy states, in the context of the BMN limit \cite{Berenstein:2002jq} and $\cal{N}$ $=4$ SYM, similar considerations apply 
\cite{Constable:2002hw}, \cite{Beisert:2002ff}, \cite{deMelloKoch:2003pv}. 
A plane-wave matrix theory \cite{Kim:2003rza} is related to the $\cal{N}$  $=4$ SYM dilatation operator  
\cite{Beisert:2004ry}. Recently, multi-matrix, multi-trace operators with diagonal free two point functions have been identified 
\cite{Kimura:2007wy}, \cite{Bhattacharyya:2008rb}. In earlier works \cite{Eguchi:1982nm}, \cite{Bhanot:1982sh}, \cite{Gross:1982at} 
it has been argued that $QCD$ can be reduced to a finite number of matrices with quenched momenta.

As such, the study of the quantum mechanics of two hermitean matrices, or equivalently of a single complex matrix, is of particular interest, as this system already possesses many of the complexities of multi-matrix models. 

Previous related studies of two hermitean matrices have typically treated the two matrices asymmetrically. For instance, in the approach first developed in \cite{Donos:2005vm}, one of the matrices, which generates the large $N$ planar background, is treated in a coordinate representation, and the other in a creation/anihilation basis. This was done in the context of $1/2$ BPS states and a dual free harmonic Hamiltonian 
\cite{Corley:2001zk}, \cite{Berenstein:2004kk}
with the matrix generating the background being the holomorphic component of a complex matrix. A precise phase space identification between the collective density description of the dynamics of this matrix \cite{Jevicki:1979mb}, and the droplet description of the LLM \cite{Lin:2004nb} metric is obtained \cite{Donos:2005vm}.  
The generalization of this approach to include $g_{YM}$ interactions was developed in \cite{Rodrigues:2005ec}. 
By considering the planar background generated by of one of the two Hermitean matrices, further properties of the spectrum were established in \cite{Cook:2007et}.
Non supersymmetric, $g_{YM}$ dependent backgrounds have been considered in \cite{Rodrigues:2008uh}, but still within an aproach where the two matrices are treated asymmetrically.

In Sections $2-5$ of this communication, we consider a more symmetrical approach to the quantum mechanics of two hermitean matrices $X_1$ and $X_2$, with $1,2$ spatial indices (or complex matrix $Z=X_1+iX_2$), by introducing matrix valued polar coordinates. In particular, we study the kinetic energy operator, or laplacian. In Section $2$, after diagonalizing the "radial" matrix, two sets of variables are identified in terms of which it is possible to invert the metric and obtain its determinant. For both sets, there is no mixing between the radial eigenvalue differentials and the angular differentials. Both metrics depend on the eigenvalues of the radial matrix. The corresponding laplacians are derived. In Section $3$, we obtain the unique laplacian that acts on gauge invariant states, expressed in terms of the eigenvalues of the radial matrix and of a single "angular" unitary matrix.  A constraint on the set of these states is identified. In Section $4$, for potentials which depend only on the radial eigenvalues, 
we show that the radially symmetric sector is equivalent to a system of non-interacting $2+1$ dimensional "s-state" fermions. In Section $5$ we consider the density description of this system, and obtain the additional potential resulting from the radial inter-eigenvalue repulsion.  

In Section $6$, we consider an arbitrary number of complex matrices and identify in this case too, the presence of repulsive radial inter-eigenvalue interactions. We conclude with some comments in Section $7$.

\noindent
\section{Matrix polar coordinates}

\noindent
We wish to consider the quantum mechanics of two $N\times N$ hermitean matrices $X_1$ and $X_2$ with hamiltonian

\bea
        \hat H &=& - \frac{1}{2} \left(\frac{\partial}{\partial (X_1)_{ij}} \frac{\partial}{\partial (X_1)_{ji}} + 
\frac{\partial}{\partial (X_2)_{ij}} \frac{\partial}{\partial (X_2)_{ji}} \right) + V(X_1,X_2) \nonumber \\
&=& - \frac{1}{2} \nabla^2 + V(X_1,X_2)
\eea

\noindent
We introduce matrix valued polar coordinates

\be\label{ZRU}
  X_1 + i X_2 = Z = R U \quad , \quad Z^{\dagger} = U^{\dagger} R
\ee

\noindent
with $R$ hermitean and $U$ unitary. Since $R$ is hermitean, it can be diagonalized as $R=V^{\dagger} r V$, with $r$ a diagonal matrix and $V$ unitary, and we obtain  

\bea\label{ZVW}
                Z = V^{\dagger} r V U &=& V^{\dagger} r W, \quad W\equiv VU \nonumber \\ 
               \quad Z^{\dagger} = U^{\dagger} V^{\dagger} r V &=& W^{\dagger} r V
\eea

\noindent
We will refer to the parametrization in terms of $(r,V,U)$ as parametrization (I), and the parametrization in terms of $(r,V,W)$ as parametrization (II). 
The number of degrees of freedom is preserved, since these matrix coordinates are defined up to $V \to D V$, $W \to D W$ with $D$ a diagonal unitary matrix.

Introducing the anti-hermitean, Lie-algebra valued differential matrices

\be\label{antidef}
        dX \equiv V dU U^{\dagger} V^{\dagger} \quad , \quad dS \equiv dV V^{\dagger} \quad , \quad dT \equiv dW W^{\dagger}
\ee

\noindent
we obtain:

\bea\label{differ}
dZ &=& V^{\dagger} (dr + [r,dS] + r dX) VU = V^{\dagger} (dr + r dT - dS r) W \nonumber \\
dZ^{\dagger} &=& U^{\dagger} V^{\dagger} (dr + [r,dS] - dX r) V = W^{\dagger} (dr + r dS - dT r) V 
\eea

\noindent
The metric is defined from the infinitesimal length squared

\bea\label{lelem}
\tr dZ^{\dagger} dZ &=& \tr \left[ (dr)^2 + [r,dS][r,dS] - r^2 (dX)^2 + [r,dS][r,dX] \right] \nonumber \\
&=& \tr \left[ (dr)^2 - r^2 (dS)^2 - r^2 (dT)^2 + 2 r  dS  r dT \right]
\eea

\noindent
Starting with parametrization (I), we note that the infinitesimal length squared has a `local'(or "pointwise") form in terms of the double index $(i,j)$, i.e.:

\bea
\tr dZ^{\dagger} dZ &=& \sum_i dr_i^2 - \sum_{ij} (r_i-r_j)^2 dS_{ij}dS_{ji} \nonumber \\
&-& {\frac{1}{2}} \sum_{ij} (r_i-r_j)^2 \{dS_{ij}dX_{ji}+dX_{ij}dS_{ji}\} \\
&-& {\frac{1}{2}} \sum_{ij} (r_i^2+r_j^2) dX_{ij}dX_{ji} \nonumber
\eea

\noindent
Recalling the antihermiticity of the differentials and writing $ds^2= g_{A,B}d\bar{x}^A dx^B$ with 
$dx^A=(dr_i,dX_{ii},dS_{{ij}(i<j)},dX_{{ij}(i<j)},dS^*_{{ij}(i<j)},dX^*_{{ij}(i<j)})$, we obtain for $g_{A,B}$:

$$
\left( 
\begin{array}{cccccc}
1 & 0 & 0 & 0 & 0 & 0 \\
0 & r_i^2 & 0 & 0 & 0 & 0 \\
0 & 0 & 0 & (r_i-r_j)^2 & {\frac{1}{2}} (r_i-r_j)^2 & 0 \\
0 & 0 & 0 & {\frac{1}{2}} (r_i-r_j)^2 & {\frac{1}{2}} (r_i^2+r_j^2) & 0 \\
0 & 0 & 0 & 0 & (r_i-r_j)^2 & {\frac{1}{2}} (r_i-r_j)^2 \\
0 & 0 & 0 & 0 & {\frac{1}{2}} (r_i-r_j)^2 & {\frac{1}{2}} (r_i^2+r_j^2)
\end{array}
\right).
$$

\noindent
It follows that 

\be\label{Vol}
   \det {g_{A,B}} = \prod_i r_i^2 \left(   \prod_{i<j} \frac{1}{4} (r_i^2-r_j^2)^2 \right)^2 = ( \prod_i r_i^2 ) (\Delta^2_{MR})^2 
\ee 

\noindent
where we have defined

\be
\Delta^2_{MR}(r_i) \equiv \prod_{i<j} \frac{1}{4} (r_i^2-r_j^2)^2  . 
\ee

\noindent
The laplacian is obtained in the standard way, and it takes the form:

\bea\label{lapone}
\nabla^2_{(I)} &=& \frac{1}{\prod_k r_k} \frac{1}{\Delta^2_{MR}} \frac{\partial}{\partial r_i}\left( {\prod_k r_k} {\Delta^2_{MR}} \right) \frac{\partial}{\partial r_i} \\ 
&+& \sum_{i\ne j} \frac{2(r_i^2+r_j^2)}{(r_i^2-r_j^2)^2} \frac{\partial}{\partial S_{ij}} \frac{\partial}{\partial S^*_{ij}} 
+ \sum_{ij} \frac{4}{(r_i+r_j)^2} \frac{\partial}{\partial X_{ij}} \frac{\partial}{\partial X^*_{ij}} \nonumber \\
&-& \sum_{i\ne j} \frac{2}{(r_i+r_j)^2} \left( \frac{\partial}{\partial S_{ij}} \frac{\partial}{\partial X^*_{ij}} + 
\frac{\partial}{\partial X_{ij}} \frac{\partial}{\partial S^*_{ij}} \right) 
 \nonumber
\eea

\noindent
For parametrization (II), we note that the second expression for the infinitesimal length squared in equation (\ref{lelem}) can be further diagonalized, for each $i$ and $j$,by introducing:

$$
dY^+=\frac{1}{\sqrt{2}} (dT+dS) \qquad dY^-=\frac{1}{\sqrt{2}} (dT-dS)
$$ 

Then

\be\label{eltwo}
\tr dZ^{\dagger} dZ = \tr [ dr^2 + \frac{1}{2} [r,dY^+][r,dY^+] - \frac{1}{2} \{r,dY^-\}\{r,dY^-\} ]
\ee

Writing again $ds^2= \tr dZ^{\dagger} dZ = g_{A,B}dx^A d\bar{x}^B$, we obtain

$$
    \det {g_{A,B}} = \prod_i 2 r_i^2  (\Delta^2_{MR})^2
$$

and straightfowardly

\bea\label{laptwo}
\nabla^2_{(II)} &=& \frac{1}{\prod_k r_k} \frac{1}{\Delta^2_{MR}} \frac{\partial}{\partial r_i}\left( {\prod_k r_k} {\Delta^2_{MR}} \right) \frac{\partial}{\partial r_i} \\ 
&+&\sum_i \frac{1}{2 r_i^2} \frac{\partial}{\partial Y^-_{ii}} \frac{\partial}{\partial Y^{*-}_{ii}}  
+ \sum_{i\ne j} \frac{2}{(r_i+r_j)^2} \frac{\partial}{\partial Y^-_{ij}} \frac{\partial}{\partial Y^{*-}_{ij}} \nonumber \\
&+& \sum_{i\ne j} \frac{2}{(r_i-r_j)^2} \frac{\partial}{\partial Y^+_{ij}} \frac{\partial}{\partial Y^{*+}_{ij}} \nonumber
\eea

\section{Invariant states}

We are interested in the action of the above laplacians, or Hamiltonian, on invariant states, i.e., states obtained by tracing strings of $Z$'s and $Z^{\dagger}$'s :

$$
            \tr (...Z^{n_p}{Z^{\dagger}}^{m_p}...Z^{n_q}{Z^{\dagger}}^{m_q}...)
$$

\noindent
These states depend only on the eigenvalues $r_i$ of the radial matrix $R$ and on the unitary matrix 

$$
                Q \equiv VUV^{\dagger} = W V^{\dagger}
$$

\noindent
Starting with parametrization (II), we note that 

$$
      dQ=dTQ-QdS= \frac{1}{\sqrt{2}}(dY^{+} Q-Q dY^{+}) + \frac{1}{\sqrt{2}}(dY^{-} Q+Q dY^{-})
$$

\noindent
from which, on this invariant subspace, 

\bea
\frac{\partial}{\partial Y^{+}_{ij}}  &=& \frac{\partial Q_{ab}}{\partial Y^{+}_{ij}}\frac{\partial}{\partial Q_{ab}} 
= \frac{1}{\sqrt{2}} (E^{(L)}_{ji}-E^{(R)}_{ji}) \nonumber \\
\frac{\partial}{\partial Y^{-}_{ij}} &=& \frac{1}{\sqrt{2}} (E^{(L)}_{ji}+E^{(R)}_{ji})
\eea

\noindent
where we have introduced the generators of left and right $U(N)$ "rotations":

$$
      E^{(L)}_{ji} = Q_{jb} \frac{\partial}{\partial Q_{ib}} \quad  E^{(R)}_{ji} = Q_{ai} \frac{\partial}{\partial Q_{aj}}            
$$

\noindent
Substitution of these expressions into the laplacian (\ref{laptwo}) yields

\bea
\nabla^2 &=& \frac{1}{\Delta^2_{MR}} \sum_i \frac{1}{r_i}  \frac{\partial}{\partial r_i}\left( r_i {\Delta^2_{MR}} \right) 
\frac{\partial}{\partial r_i}  
- \sum_i \frac{1}{4 r_i^2} (E^{(L)}_{ii}+E^{(R)}_{ii})(E^{(L)}_{ii}+E^{(R)}_{ii})  \\ 
&-&\sum_{i \ne j} \left( \frac{2(r_i^2+r_j^2)}{(r_i^2-r_j^2)^2} (E^{(L)}_{ij}E^{(L)}_{ji}+E^{(R)}_{ij}E^{(R)}_{ji})
 - \frac{4 r_i r_j}{(r_i^2-r_j^2)^2} (E^{(L)}_{ij}E^{(R)}_{ji}+E^{(R)}_{ij}E^{(L)}_{ji}) \right) \nonumber
\eea

\noindent
Before arriving at the final expression, note from (\ref{eltwo}) that $dY^{+}_{ii}$ is absent, and therefore we require:

\be\label{const}
0 = \frac{\partial}{\partial Y^{+}_{ii}} = \frac{1}{\sqrt{2}} (E^{(L)}_{ii}-E^{(R)}_{ii}) \quad ; \quad E^{(L)}_{ii}=E^{(R)}_{ii}
\ee

\noindent
The final form of the laplacian is then:

\bea\label{lapinv}
\nabla^2 &=& \frac{1}{\Delta^2_{MR}} \sum_i \frac{1}{r_i}  \frac{\partial}{\partial r_i}\left( r_i {\Delta^2_{MR}} \right) 
\frac{\partial}{\partial r_i}  
- \sum_i \frac{1}{r_i^2} E^{(L)}_{ii} E^{(L)}_{ii} \\ 
&-&\sum_{i \ne j} \left( \frac{2(r_i^2+r_j^2)}{(r_i^2-r_j^2)^2} (E^{(L)}_{ij}E^{(L)}_{ji}+E^{(R)}_{ij}E^{(R)}_{ji})
 - \frac{4 r_i r_j}{(r_i^2-r_j^2)^2} (E^{(L)}_{ij}E^{(R)}_{ji}+E^{(R)}_{ij}E^{(L)}_{ji}) \right) \nonumber
\eea

\noindent
In terms of parametrization (I), from

$$
dP \equiv dQ Q^{\dagger} = dS + dX - Q dS Q^{\dagger}
$$

\noindent 
one obtains, when acting on invariant states, 

\bea
\frac{\partial}{\partial X_{ij}}  = E^{(L)}_{ji} \qquad \frac{\partial}{\partial S_{ij}} = E^{(L)}_{ji}-E^{(R)}_{ji} .
\eea

\noindent
Substitution of these in (\ref{lapone}) results in the laplacian (\ref{lapinv}), with states having to satisfy the contstraint (\ref{const}),
now resulting from the $dS_{ii}$ mode. As expected, the form of the laplacian (\ref{lapinv}) is unique.

\section{Radial fermionization}

In the case of the simple integral ($d=0$ system) over two hermitean matrices, it follows from the form of the "`volume"' element (\ref{Vol}) that an additional term is added to the action expressing the the repulsive  inter-egenvalue interaction amongst the radial eigenvalues. By either integrating out the angular dependence, or by factoring it out if the action is only dependent on the radial eigenvalues (as is the case for the free theory in an harmonic potential) \cite{Mtho}, one straighforwardly obtains a BIPZ  \cite{Brezin:1977sv} type equation. 

For the hamiltonian system, it is a well known result \cite{Brezin:1977sv} that the singlet sector of a $N \times N$ single hermitean matrix hamiltonian with a potential depending only on 
its eigenvalues is equivalent to a system of $N$ non-interacting fermions. This is a result of the anti-symmetry under the exchange of any two coordinates of the Van der Monde determinant

$$
\Delta (x_k) = \prod_{i<j} (x_i-x_j)
$$

We consider the case of a potential that depends only on the eigenvalues of the radial matrix $U$. An example would a potential of the form:

$$
V(X_1,X_2) = \tr v(ZZ^{\dagger}) = \tr v(Z^{\dagger}Z)  
$$

\noindent
with $v(x)$ a polynomial. Then, on "s-states" (independent of the "angular" degrees of freedom $Q$), and letting $\rho_i=r_i^2$,

$$
- \frac{1}{2} \nabla^2 = - \frac{1}{2} \frac{1}{\Delta^2_{MR}} \sum_i \frac{1}{r_i}  \frac{\partial}{\partial r_i}\left( r_i {\Delta^2_{MR}} \right)
\frac{\partial}{\partial r_i}
 = - \frac{2}{\Delta^2(\rho)} \sum_i  \frac{\partial}{\partial \rho_i} \left( \rho_i {\Delta^2(\rho)} \right)\frac{\partial}{\partial \rho_i}
$$

\noindent
This kinetic energy operator acts on symmetric wavefunctions $\Phi$. Defining 

\be\label{wav}
                                        \Psi = \Delta~\Phi \quad ,
\ee

\noindent
its action on $\Psi$ takes the form:

\bea
&-& \frac{2}{\Delta} \sum_i  \frac{\partial}{\partial \rho_i}  \rho_i {\Delta^2} \frac{\partial}{\partial \rho_i} \frac{1}{\Delta} = - 2 \sum_i \frac{1}{\Delta} \frac{\partial}{\partial \rho_i} \Delta ~\rho_i ~\Delta  \frac{\partial}{\partial \rho_i} \frac{1}{\Delta} \nonumber \\
= &-& 2 \sum_i \left(    \frac{\partial}{\partial \rho_i} + \sum_{k\ne i} \frac {1}{\rho_i-\rho_k}\right) \rho_i
\left(    \frac{\partial}{\partial \rho_i} - \sum_{j\ne i} \frac {1}{\rho_i-\rho_j}\right) \\
= &-& 2 \sum_i \left( \frac{\partial}{\partial \rho_i} \rho_i  \frac{\partial}{\partial \rho_i}-\sum_{j\ne i} \frac {1}{\rho_i-\rho_j}
+  \sum_{j\ne i} \frac {\rho_i}{(\rho_i-\rho_j)^2} - \sum_{j\ne i}\sum_{k\ne i} \frac {\rho_i}{\rho_i-\rho_k}\frac {1}{\rho_i-\rho_j} \right)
\nonumber
\eea

\noindent
The second term clearly vanishes, and the last two vanish due to the identity

$$
 \sum_{\begin{array}{c} i,j,k\\ i \ne j \\ i \ne k \\ j \ne k \end{array}} \frac {\rho_i}{(\rho_i-\rho_k)(\rho_i-\rho_j)}     = 0
$$

\noindent
which generalizes the identity applicable to the single hermitean case. It is easily proven by choosing any three distinct eigenvalues.

Therefore, the eigenvalue equation for the energies of the system takes the form, 

\be\label{lapfree}
\left( - 2 \sum_i  \frac{\partial}{\partial \rho_i} \rho_i  \frac{\partial}{\partial \rho_i} + v(\rho_i) \right) \Psi =
\left( - \frac{1}{2} \sum_i \frac{1}{r_i} \frac{\partial}{\partial r_i} r_i  \frac{\partial}{\partial r_i} + v'(r_i) \right) \Psi 
= E \Psi
\ee

\noindent
This is the "s-state" Schroedinger equation for $N$ non-interacting $2+1$ dimensional non-relativistic fermions. 

\section{Density description}

In this section we describe the key features of the density descripton of radially symmetric fermions int terms of the density of radial eigenvalues. We use the collective field theory approach of Jevicki and Sakita \cite{Jevicki:1979mb}. 

The existence of such a descripton requires the identification of a suitable set of invariant operators which close under "`joining"' and "`splitting"', equivalent to the closure of underlying Schwinger-Dyson equations.

Remarkably, the following set can be identified as such:

\be\label{}
\Phi_k = \tr e^{ik Z^{\dagger}Z} = \sum_i e^{ik r_i^2} \quad; 
\quad \Phi(x)= \int \frac{dk}{2\pi}e^{-ikx} \Phi_k = \sum_i \delta(x-r_i^2) 
\ee      

\noindent
One has

\be
\frac{\partial \Phi_k }{\partial Z^{\dagger}_{ij}} = ik \left( Z e^{ik Z^{\dagger}Z}  \right)_{ji} ~, \quad  
\frac{\partial \Phi_k }{\partial Z_{ij}} = ik \left( e^{ik Z^{\dagger}Z} Z^{\dagger} \right)_{ji} 
\ee

\noindent
from which the "`joining"' operator $\Omega$ \cite{Jevicki:1979mb} takes the form

\bea\label{bw}
\Omega_{kk'} &=& \frac{\partial \Phi_k }{\partial Z^{\dagger}_{ij}}  \frac{\partial \Phi_{k'}}{\partial Z_{ji}}  =
- k k' \tr Z^{\dagger} Z e^{i(k+k') Z^{\dagger}Z}  \\
\Omega_{xx'}&=& \int \frac{dk}{2\pi} \int \frac{dk'}{2\pi} e^{-ikx} e^{-ik'x} \Omega_{kk'} = 
\partial_x \partial_{x'} \left[ x \Phi(x) \delta (x-x')    \right] \nonumber
\eea

For the "`splitting"' operator $\omega$ \cite{Jevicki:1979mb} 

\bea
\omega_k &=& \frac{\partial^2 \Phi_k }{\partial Z^{\dagger}_{ij}\partial Z_{ji}} = - k \int_0^k dk' \Phi_k \tr Z^{\dagger} Z e^{i(k-k') Z^{\dagger}Z}+ i k N \Phi_k \nonumber \\
&=& - k \sum_{ij} \int_0^k dk' e^{i k' r_i^2} e^{i (k-k') r_j^2}r_j^2 + ikN \sum_i e^{i k' r_i^2}
\eea

\noindent
This term requires careful manipulation:

\bea
\omega_k &=& - k \sum_{ij} \int_0^k dk' e^{i k' r_i^2} e^{i (k-k') r_j^2}r_j^2 + ikN \sum_i e^{i k r_i^2} \\
&=& - k \sum_{i\ne j} \int_0^k dk' e^{i k' r_i^2} e^{i (k-k') r_j^2}r_j^2 - k^2 \sum_i r_i^2 e^{i k r_i^2}
+ ikN \sum_i e^{i k r_i^2} \nonumber \\
&=& ik \sum_{i\ne j} r_j^2  \frac{ e^{i k r_i^2}- e^{i k r_j^2}}{r_i^2 - r_j^2}- k^2 \sum_i r_i^2 e^{i k r_i^2}
+ ikN \sum_i e^{i k r_i^2} \nonumber \\
&=& -2ik \sum_{i\ne j}   \frac{ r_j^2 e^{i k r_j^2}}{r_i^2 - r_j^2}-ik (N-1) \sum_i e^{i k r_i^2} - k^2 \sum_i r_i^2 e^{i k r_i^2}
+ ikN \sum_i e^{i k r_i^2} \nonumber\\
&=& -2ik \sum_{i\ne j}   \frac{ r_j^2 e^{i k r_j^2}}{r_i^2 - r_j^2}- k^2 \sum_i r_i^2 e^{i k r_i^2}
+ ik \sum_i e^{i k r_i^2} \nonumber
\eea

\noindent
hence

\bea
w_x&=&\int \frac{dk}{2\pi} e^{-ikx} w_k =
-\partial_x \left[ 2 x \Phi(x) \int \frac{dy \Phi(y)}{x-y} + \Phi(x) - \partial_x (x \Phi(x))\right] \nonumber \\
&=& -\partial_x \left[ x \Phi(x) \left( 2 \int \frac{dy \Phi(y)}{x-y} - \frac{\partial_x \Phi(x)}{\Phi(x)}\right) \right] 
\eea

\noindent
As it has been shown to be consistent \cite{DasKA}, 
the $\partial_x \Phi(x)/\Phi(x)$ term can be neglected
\footnote{It is associated with a $i=j$ regularization and is, in any case, subleading in $1/N$.} and we have finally 

\be\label{lw}
w_x= -\partial_x \left[ x \Phi(x) \left( 2 \int \frac{dy \Phi(y)}{x-y}\right)\right]
\ee 

\noindent
In the density variable descripton \cite{Jevicki:1979mb}, the Jacobian $J$ associated with the change of variables to the invariant operators satisfies:

$$
\left( \int dx' \Omega_{xx'} \frac{\partial}{\partial \Phi(x')} + w_x + 
\int dx' \frac{\partial \Omega_{xx'}}{\partial \Phi(x')} \right) ~J = 0  
$$

\noindent
Since $\int dx' \partial \Omega_{xx'} / \partial \Phi(x')=0$, if follows from (\ref{bw}) and (\ref{lw}) that

$$
\partial_x \frac{\partial}{\partial \Phi(x)} ~\ln J = 2 \int \frac{dy \Phi(y)}{x-y}
$$

\noindent
The solution is

$$
ln J = \int dx \int dy \Phi(x) \Phi(y) \ln |x-y| ~; \quad   J = \prod_{i<j} \frac{1}{4}(r_i^2-r_j^2)^2 = \Delta^2_{MR}(r_i)
$$    
    
\noindent
in precise agreement with the results of Section $2$, e.g., (\ref{Vol}). The prefactor in (\ref{Vol}) is simply the result of the change of variables $x=r^2$.
    
How does the repulsion amongst the radial eigenvalues express itself as a contribution to the potential? This is given by  \cite{Jevicki:1979mb}:

$$
\omega \Omega^{-1} \omega = \int_0^{\infty} dx ~x \Phi(x) 
\left( \int_0^{\infty} \frac{dy ~\Phi(y)}{x-y} \right)^2
$$ 

\noindent
Let us introduce a density of radial eigenvalues $\phi(r)$ such that

$$
 \int_0^{\infty} dx \Phi(x) ~f(x) =   \int_0^{\infty} 2 r dr \Phi(r^2) ~f(r^2) \equiv 
 \int_0^{\infty} dr \phi(r) ~f(r^2)
$$ 

\noindent
and extend the domain of definition of $\phi(r)$ to the real line by requiring $\phi(-r)=\phi(r)$. Then

\bea
\omega \Omega^{-1} \omega &=& \int_0^{\infty} dr ~r^2 \phi(r) 
\left( \int_0^{\infty} \frac{ds ~\phi(s)}{r^2-s^2} \right)^2 \nonumber \\
&=& \frac{1}{2} \int_{-\infty}^{\infty} dr  \phi(r) 
\left( \int_0^{\infty} \frac{ds ~r ~\phi(s)}{r^2-s^2} \right)^2 \nonumber \\
&=& \frac{1}{8} \int_{-\infty}^{\infty} dr  \phi(r) 
\left( \int_{-\infty}^{\infty} \frac{ds ~\phi(s)}{r-s} \right)^2 \nonumber \\
&=&\frac{\pi^2}{24} \int_{-\infty}^{\infty} dr  \phi^3(r)  
\eea

\noindent
Remarkably, as is the case in the collective field descripton of the singlet sector of the single hermitean matrix
\footnote{In this case the cubic potential is the Thomas-Fermi density term of $1$ dimensional fermions}, a local cubic potential is generated in the bosonized radially symmetric sector of the $2+1$ fermions.

\section{More complex matrices}

As is well known, the case of larger number of matrices is of great importance in the context of the AdS/CFT corespondence  \cite{Maldacena:1997re}, \cite{Gubser:1998bc}, \cite{Witten:1998qj}. Of particular importance is the case of $3$ complex matrices, which are associated with the $3$ Higgs of the bosonic sector of $\textit{N}=4$ SYM.

Let us consider in general $m$ complex matrices $Z_A ~, A=1,...,m$. Then 

$$
          \sum_A Z_A^{\dagger}Z_A
$$

\noindent 
is an Hermitean, positive definite matrix. As in the previous subsections, we will denote its eigenvalues by $r_i^2$. The corresponding density variables are:

\be
\Phi_k = \tr e^{ik \sum_B Z_B^{\dagger}Z_B} = \sum_i e^{ik r_i^2} \quad; 
\quad \Phi(x)= \int \frac{dk}{2\pi}e^{-ikx} \Phi_k = \sum_i \delta(x-r_i^2) 
\ee    

\noindent
One has

\be
\frac{\partial \Phi_k }{\partial (Z_A)^{\dagger}_{ij}} = ik \left( Z_A e^{ik Z_B^{\dagger}Z_B}  \right)_{ji} \qquad 
\frac{\partial \Phi_k }{\partial (Z_A)_{ij}} = ik \left( e^{ik Z_B^{\dagger}Z_B} Z_A^{\dagger} \right)_{ji} 
\ee

\noindent 
For $\Omega$, the result is identical to the that of the previous section:

\bea\label{bwm}
\Omega_{kk'} &=& \frac{\partial \Phi_k }{\partial (Z_A)^{\dagger}_{ij}}  \frac{\partial \Phi_{k'}}{\partial (Z_A)_{ji}}  = - k k' \tr Z_A^{\dagger} Z_A e^{i(k+k') Z_B^{\dagger}Z_B}  \\
\Omega_{xx'}&=& \int \frac{dk}{2\pi} \int \frac{dk'}{2\pi} e^{-ikx} e^{-ik'x} \Omega_{kk'} = 
\partial_x \partial_{x'} \left[ x \Phi(x) \delta (x-x')    \right] \nonumber
\eea

\noindent
For $\omega$, one obtains

\bea
\omega_k &=& \frac{\partial^2 \Phi_k }{\partial (Z_A)^{\dagger}_{ij}\partial (Z_A)_{ji}} = - k \int_0^k dk' \Phi_k \tr Z^{\dagger} Z e^{i(k-k') Z^{\dagger}Z}+ i k ~m N ~\Phi_k \nonumber \\
&=& - k \sum_{ij} \int_0^k dk' e^{i k' r_i^2} e^{i (k-k') r_j^2}r_j^2 + i k ~m N ~\sum_i e^{i k' r_i^2}
\eea

\noindent
As described in the previous section, this yields

\be\label{lwm}
w_x= -\partial_x \left[ x \Phi(x) \left( 2 \int \frac{dy \Phi(y)}{x-y} + \frac{N(m-1)}{x}\right)\right]
\ee 

\noindent
and hence the Jacobian satisfies

$$
\partial_x \frac{\partial}{\partial \Phi(x)} ~\ln J = 2 \int \frac{dy \Phi(y)}{x-y} + \frac{N(m-1)}{x}
$$

\noindent
A full treatment of the ensuing Jacobian is beyond the scope of this article, but the first term on the right hand of side shows it to be multiplied by the Van der Monde type determinant $\Delta^2_{MR}$, associated with the radial inter-eigenvalue repulsion.  

\section{Comments and Conclusion}

By introducing matrix valued polar coordinates, we were able to obtain the form of a two hermitean matrix hamiltonian in terms of the eigenvalues of the ``radial matrix" (or the eigenvalues of $Z Z^{\dagger}$)  and of a single ``angular" unitary matrix ($Q$). For potentials which only depend on the radial eigenvalues, the radial Schroedinger equation can be mapped to the Schroedinger equation of a system of non-interacting $2+1$ dimensional "`radial fermions"'. We obtained the form of the additional potential resulting from the repulsive interaction amongst the radial eigenvalues in the density description of the system. The presence of repulsive interactions amongst the radial eigenvalues has been identified for an arbitrary number of complex matrices. 

The emergence of higher dimensional laplacians from matrices and the appearance, as a result of quantum effects, of repulsive inter-eigenvalue interactions, is potentially of great interest to the AdS/CFT correspondence \cite{Maldacena:1997re}, \cite{Gubser:1998bc}, \cite{Witten:1998qj} and to the general issue of the emergence of gravity (e.g.,\cite{Berenstein:2005aa} ) and it deserves further study.

As a note of clarification, clearly the use of the term "fermionization" is appropriate to the one dimensional radial wave funtions described in this communication. We do not use it in the sense of a full $2+1$ dimensional fermionization. However, even in the presence of "angular" excitations, the redefinition (\ref{wav}) would still map the laplacian (\ref{lapinv}) to the much simpler form of (\ref{lapfree}), added by the same angular contribution. 

\section{Acknowledgements}

We would like to thank Antal Jevicki and Robert de Mello Koch for comments on an earlier version of this article. J.P.R. would like to thank the hospitality extended to him by the Theory High Energy Group of Brown University during a recent visit, when most of this article was written up.

\end{document}